\documentclass[letterpaper,twocolumn,aps,pra,showpacs,superscriptaddress]{revtex4}
\usepackage{ulem}
\usepackage{amsmath,mathtools,amssymb}
\usepackage{graphicx}
\usepackage{sidecap}
\usepackage{dcolumn}
\usepackage{color}\pagecolor{white}
\usepackage{bm}
\usepackage{hyperref}

\begin{document}


\title{Internal Decoherence of a Gaussian Wave Packet in a Harmonic Potential}

\author{A. Cidrim}
\affiliation{Instituto de F\'{i}sica de S\~{a}o Carlos, Universidade de S\~{a}o Paulo, Caixa Postal 369, 13560-970, S\~{a}o Carlos, SP, Brazil}
\author{F. E. A. dos Santos}
\affiliation{Instituto de F\'{i}sica de S\~{a}o Carlos, Universidade de S\~{a}o Paulo, Caixa Postal 369, 13560-970, S\~{a}o Carlos, SP, Brazil}
\affiliation{Departamento de F\'{i}sica, Universidade Federal de S\~{a}o Carlos, 13565-905, S\~{a}o Carlos, SP, Brazil}
\author{A. O. Caldeira} %
\affiliation{Instituto de F\'{i}sica Gleb Wataghin, Universidade Estadual de Campinas, 13083-859, Campinas, SP, Brazil}




\date{\today}

\begin{abstract}
We have studied the quantum dissipative problem of a Gaussian wave packet under
the influence of a harmonic potential. A phenomenological approach to dissipation is adopted in
the light of the well-known model in which the environment is composed of a bath
of non-interacting harmonic oscillators.
As one of the effects of the coupling to the bath is the evolution of an initially pure wave packet into
a statistical mixture, we estimate the characteristic time elapsed for this to occur for different
regimes of temperature, damping, and also different initial states.
\end{abstract}

\maketitle


\section{Introduction}

Nowadays, the importance of studying dissipative quantum systems is indisputable, in view of its practical impact on promising new technologies, envisaged by the  fields of quantum computation and information, and also of its role in discussions about fundamentals of quantum mechanics. For instance,  regarding the environment as a selector of a preferential basis has been widely explored in several works and considered as a support  for the arguments in favor of the decoherence interpretation of the measurement problem \cite{Zurek2003a}. Some of these works suggest that the emergence of classicality in a quantum system would be intimately related to the phenomenon of decoherence induced by its interaction with practically infinite degrees of freedom. This latter topic is by itself a prolific source of discussion in the scientific community and still leaves some open questions \cite{Leggett2002}.

Motivated by these exciting perspectives, we study a simple dissipative quantum system, composed of a single Gaussian wave packet in a harmonic potential. The environment is described by a bath of non-interacting harmonic oscillators coupled to the system of interest through a coordinate-coordinate interaction and we explicitly evaluate the time evolution of the reduced density operator of the latter. In the coordinate representation, this density operator can be viewed as a matrix represented in a continuous basis, and whose off-diagonal elements are a measure of the internal coherence of the packet. As time evolves, there is a tendency to the diagonalization of the matrix as an effect of the interaction with the bath.

Actually, the first time the above-mentioned system-plus-environment model was employed to investigate the loss of coherence in a quantum mechanical system  was in \cite{Caldeira1985} where the time evolution of the interference between two Gaussian wave packets in a harmonic potential was studied. It was shown that the interference term relaxes within a very short time scale given by the natural relaxation time of the system divided by the square of the initial distance between the centers of the packets measured in units of their initial widths. In other words, the further they initially are the faster will interference   disappear. However, nothing was said about the coherence still left within each wave packet which ultimately contributes to the purity of the superposition state under investigation.

In this paper, our goal is to estimate the time elapsed for the vanishing of the off-diagonal elements of the reduced density operator corresponding to a single Gaussian wave packet, in this way, defining what we shall call the \textit{internal decoherence time} of the packet. Our analysis differs from what is typically found in the literature, once we make no use of approximated master equations. We derive our expressions rigorously from the reduced density operator of the system and extract this internal decoherence time from the resulting analytical expressions in several regimes of temperature and dissipation, which present characteristic times that can be important for some particular physical systems.

This paper is structured as follows: in Sec.\ref{model} we describe the derivation of the reduced density operator for the Gaussian wave packet using the well-known approach developed in \cite{Caldeira1983} for dissipative quantum systems. Next, a general discussion of the time scales  involved in the problem is held in Sec. \ref{time}. Then, our results for both weak and strong damping (Sec.\ref{subweak} and \ref{substrong}) are analyzed, for different limits of temperature. Finally a brief discussion of our results is then presented in Sec.\ref{discussion}.
 
\section{ Gaussian wave packet and the dissipative model}\label{model}
Suppose that our system is given by a particle of mass $M$ in a harmonic potential with characteristic frequency $\omega_0$ and whose width of the ground state wave function squared is $ \sigma_0^2\equiv \hbar/2M\omega_{0}$. We use a system-plus-reservoir model to describe our problem, in which the reservoir is composed by a bath of noninteracting harmonic oscillators which are bilinearly coupled in coordinate to the system. 
Making use of the Feynman-Vernon approach we can describe the system by its reduced density operator, after the information concerning the bath degrees of freedom is traced out. Its coordinate representation is given by 
\begin{equation}\label{rhored}
\rho(x,y,t)=\int\int dx'dy' J(x,y,t;x',y',0)\rho(x',y',0),
\end{equation}
where the primed coordinates refer to the initial position at the time $t'=0$ and 

\begin{widetext}

\begin{equation}\label{superJ}
\begin{aligned}
J(x,y,t;x',y',0)&=\int_{x'}^{x}\int_{y'}^{y}\mathcal{D}x(t')\mathcal{D}y(t')
\exp\frac{i}{\hbar}\Big\{S_{0}[x(t')]-S_{0}[y(t')]-M\gamma\int_{0}^{t}(x\dot{x}-y\dot{y}+x\dot{y}-y\dot{x})dt'\Big\}\\
&\times\exp-\Big\{\frac{2M\gamma}{\pi\hbar}\int_{0}^{\Omega}d\nu \nu \coth\left(\frac{\hbar\nu}{2k_{B}T}\right)
\int_{0}^{t}\int_{0}^{\tau}d\tau d\sigma [x(\tau)-y(\tau)]\cos[{\nu(\tau-\sigma)}][x(\sigma)-y(\sigma)]\Big\},
\end{aligned}
\end{equation}

\end{widetext}
is the superpropagator of the model, which carries the initial state from $(x',y',0)\rightarrow (x,y,t)$. The  functional $S_{0}[x(t')]$ is the classical action of an arbitrary path $x(t')$ of the particle in the harmonic potential. The parameter $\gamma\equiv\eta/2M$ is the  relaxation frequency of the particle and $\eta$ the dissipative constant from the classical (Markovian) Langevin equation, $\Omega$ is a high-frequency cut-off associated with the typical microscopic short times involved in the problem. At last, the temperature $T$ is assumed to be the one at which the oscillators of the bath are in thermal equilibrium.  

The frequency that characterizes whether we have under or overdamped motion is given by
\begin{equation}\label{freq}
\omega^2\equiv \omega_{0}^2-\gamma^2,
\end{equation}
which together with the newly defined dimensionless variables  
\begin{equation}
\begin{aligned}
&R\equiv\frac{\gamma}{\omega_{0}},\quad S\equiv\frac{\omega}{\omega_{0}},\quad \kappa\equiv\frac{\hbar\omega_0}{2k_{B}T},\\
&\theta\equiv \omega_{0}t,\quad \lambda\equiv\frac{\nu}{\omega_0},
\end{aligned}
\end{equation}
will play an important role in our forthcoming developments. It is easy to see that (\ref{freq}) can be rewritten in terms of these variables, giving the useful relation
\begin{equation}
R^2+S^2=1.
\end{equation}

Assuming that the initial state of the particle is a pure Gaussian state, we can write
\begin{equation}
\rho(x',y',0)=\frac{1}{\sqrt{2\pi\sigma^2}}\exp{i\frac{p(x'-y')}{\hbar}}\exp{\frac{(x'^2+y'^2)}{4\sigma^2}},
\end{equation}
where the initial preparation of the state is evidenced by its width $\sigma$ and momentum $p$.

It is also interesting to define new dimensionless variables in terms of $\sigma_0$ as  
\begin{equation}
q\equiv\frac{x+y}{2\sigma_0};\qquad
r\equiv \frac{x-y}{\sigma_0},
\end{equation}
where $r$ has the important meaning of how distant the matrix element is from the principal diagonal. 

Since we are dealing with a problem that involves only quadratic Lagrangians, we can exactly solve the path integrals in (\ref{superJ}). Those turn out to be simply Gaussian integrations, allowing us to write the reduced density operator (\ref{rhored}) as
\begin{equation}\label{roabrev}
\begin{aligned}
\rho(q,r,\theta)=&\sqrt{\frac{P(\theta)}{\pi}}\exp-\left\{P(\theta)\left(q-\frac{p}{N(\theta)}\right)^2\right\}\\
&\times\exp-F(\theta)r^2\exp {iD(q,p,\theta)r}.
\end{aligned}
\end{equation}
The functions $P(\theta)$, $N(\theta)$, and $F(\theta)$ depend only on time and the exponential term containing $D(q,p,\theta)$ will be only an oscillatory contribution to (\ref{roabrev}). Therefore, $D(q,p,\theta)$ will not be used anywhere in this work. The combination $p/N(\theta)$ is nothing but the classical trajectory of a damped particle with the appropriate initial conditions for this problem and the remaining functions will be defined below. 

As we are interested in the process of the initially pure wave packet turning into a statistical mixture, it is convenient to use a purity measurement, in our case chosen to be the trace of the squared density operator (\ref{roabrev}), which is given by
\begin{equation}\label{purity}
s=\text{Tr}\rho^2=\frac{1}{2}\sqrt{\frac{P(\theta)}{F(\theta)}}.
\end{equation} 

From (\ref{roabrev}) we can also see that the second exponential determines the way in which the off-diagonal elements of the density matrix decay. It is then clear that, for our purposes, the function $F(\theta)$ is the most important object, since it is by analyzing its time dependence that we can estimate the internal decoherence time of the wave packet. Its explicit form is given by

\begin{widetext}
\begin{equation}\label{gamma}
F(\theta)r^2\equiv\frac{1}{\sin^2(S\theta)} \left\{A^{(1)}(\theta)+\frac{S^2e^{-2R\theta}}{8}\zeta^2-\frac{1}{8}\frac{\Big[ Se^{-R\theta}K(\theta)\zeta^2-4A^{(2)}(\theta)\Big]^2}{K^2(\theta)\zeta^2+8 A^{(3)}_{1}(\theta)}\right\}r^2,
\end{equation}
with $\zeta\equiv\sigma/\sigma_0$. 

Since our initial state is Gaussian, it always satisfies the minimum of Heisenberg's uncertainty (i.e, $\Delta x\Delta p=\hbar/2$). Therefore, in this case, $\zeta$ can be regarded as a squeezing parameter, telling us how the spatial dispersion $\sigma\equiv\Delta x$ deviates from the ideal non-squeezed value $\sigma_0$. 

For the evaluation of the purity using (\ref{purity}), we must also know the explicit form of 
\begin{equation}
P(\theta)\equiv \frac{1}{4}\frac{e^{2R\theta}}{K^2(\theta)\zeta^2+8 A^{(3)}_{1}(\theta)},
\end{equation}
following the definitions: 
\begin{equation}\label{LK}
A^{(3)}_{1}(\theta)\equiv A^{(3)}(\theta)+\frac{\sin^2(S\theta)}{8\zeta^2},\qquad K(\theta)\equiv S\cos({S\theta})+R\sin(S\theta).
\end{equation}

Finally, the functions 
\begin{equation}\label{Athetas}
A^{(i)}(\theta)\equiv\frac{R}{2\pi}\int_{0}^{\lambda_{c}}d\lambda \lambda\coth{(\kappa\lambda)}A^{(i)}(\lambda,\theta),
\end{equation}
with high frequency cut-off $\lambda_C\equiv\Omega/\omega_0$, carry all the temperature dependence of the problem. The remaining functions in the above defined expressions are
\begin{equation}\label{Athetas1}
A^{(1)}(\lambda,\theta) = e^{-2R\theta}\int_{0}^{\theta}\int_{0}^{\theta}\sin{(S\theta_{1})}\cos{[\lambda(\theta_{1}-\theta_{2})]}\sin{(S\theta_{2})}\exp{[R(\theta_{1}+\theta_{2})]}d\theta_{1}d\theta_{2},
\end{equation}

\begin{equation}\label{Athetas2}
A^{(2)}(\lambda,\theta) = 2e^{-R\theta}\int_{0}^{\theta}\int_{0}^{\theta}\sin{(S\theta_{1})}\cos{[\lambda(\theta_{1}-\theta_{2})]}\sin{[S(\theta-\theta_{2})]}\exp{[R(\theta_{1}+\theta_{2})]}d\theta_{1}d\theta_{2},
\end{equation}

\begin{equation}\label{Athetas3}
A^{(3)}(\lambda,\theta) = \int_{0}^{\theta}\int_{0}^{\theta}\sin{[S(\theta-\theta_{1})]}\cos{[\lambda(\theta_{1}-\theta_{2})]}\sin{[S(\theta-\theta_{2})]}\exp{[R(\theta_{1}+\theta_{2})]}d\theta_{1}d\theta_{2}.
\end{equation}
\end{widetext}

Our evaluation of the internal decoherence time $\tau_D$ will follow the analysis of the exponent (\ref{gamma}). Generally, the dominant temporal behavior will be linear, as we will confirm below, so that we can write $-F(\theta)r^2\approx -\Gamma t+c$, where $c$ is a constant. We define $\tau_D\equiv\Gamma^{-1}$ as the typical time after which the off-diagonal terms decay to $1/e$ of their initial value. Of course, this quantity depends inversely on how far the matrix elements are from the diagonal, i.e., on the distance $r$. 

\section{Time scales}\label{time}

It is very important to separate the time scales involved in each dissipative regime. As we are interested in the behavior of the system at very long times, we look for results that do not depend on the particular cut-off frequency chosen, and therefore the limit $\lambda_C\rightarrow\infty$ (in practice $\lambda_C\gg 1$) turns out to be an appropriate approximation. Actually, we are usually considering the cut-off larger than any other relevant frequencies, namely, $\kappa^{-1}$ and $R$, associated with the thermal and dissipation frequencies, respectively.

To guarantee the validity of this hypothesis, we must analyze carefully the general behavior of the integral (\ref{Athetas}).

Beginning with the underdamped case (i.e., $R<1$), it is convenient to evaluate the double time integrals in (\ref{Athetas1}-\ref{Athetas3}) and rewrite the resulting expressions as

\begin{widetext}
\begin{equation}\label{a1-1}
\begin{aligned}
&A^{(1)}(\theta)=\frac{1}{8}\Bigg\{\left[\frac{S^2e^{-2R\theta}}{\pi}+\frac{[R\sin(S\theta)-S\cos{(S\theta)}]^2}{\pi}\right]I(0)\\
&-\frac{2Se^{-R\theta}[S\cos(S\theta)-R\sin{(S\theta)}]}{\pi}I(\theta)+\frac{2Se^{-R\theta}\sin{(S\theta)}}{\pi}\frac{dI(\theta)}{d\theta}-\frac{\sin^2{(S\theta)}}{\pi}\frac{d^2I(\theta)}{d\theta^2}\Big|_{\theta=0}\Bigg\},
\end{aligned}
\end{equation}

\begin{equation}\label{a2-1}
\begin{aligned}
&A^{(2)}(\theta)=\frac{1}{4}\Bigg\{2\left[\frac{RS \sin(S\theta)\sinh(R\theta)-S^2\cos(S\theta)\cosh(R\theta)}{\pi}\right]I(0)+\frac{2S^2-\sin^2(S\theta)}{\pi}I(\theta)\\
&-\frac{2S\sin(S\theta)\cos(S\theta)}{\pi}\frac{dI(\theta)}{d\theta}+\frac{\sin^2{(S\theta)}}{\pi}\frac{d^2I(\theta)}{d\theta^2}\Bigg\},
\end{aligned}
\end{equation}

\begin{equation}\label{a3-1}
\begin{aligned}
&A^{(3)}(\theta)=\frac{1}{8}\Bigg\{\left[\frac{S^2e^{2R\theta}}{\pi}+\frac{[R\sin(S\theta)+S\cos{(S\theta)}]^2}{\pi}\right]I(0)\\
&-\frac{2Se^{R\theta}[S\cos(S\theta)+R\sin{(S\theta)}]}{\pi}I(\theta)+\frac{2Se^{R\theta}\sin{(S\theta)}}{\pi}\frac{dI(\theta)}{d\theta}-\frac{\sin^2{(S\theta)}}{\pi}\frac{d^2I(\theta)}{d\theta^2}\Big|_{\theta=0}\Bigg\},
\end{aligned}
\end{equation}
with the definition of the integral
\begin{equation}\label{inttheta2-1}
I(\theta)\equiv\frac{1}{2}\int_{-\lambda_{c}}^{\lambda_{c}}d\lambda\frac{4R\lambda}{(\lambda^2-1)^2+4R^2\lambda^2}\coth(\kappa\lambda)\cos(\lambda\theta),
\end{equation}
where $\lambda_C\gg 1$.
\end{widetext}

It is possible to sketch the general behavior of the function $F(\theta)$ by taking its asymptotic limits. It can be shown, substituting equations (\ref{a1-1}-\ref{a3-1}) into (\ref{gamma}), that 
\begin{equation}
F(\theta)=\begin{dcases}
\frac{1}{8\zeta^2}, & \mbox{for } \theta\rightarrow 0,\\
-\frac{1}{8\pi}\frac{d^2I(\theta)}{d\theta^2}\Big|_{\theta=0}, & \mbox{for } \theta\rightarrow \infty,
\end{dcases}
\end{equation}
where in the last case it was considered that the contributions from the integral (\ref{inttheta2-1}) and its derivatives go to zero in the limit of long times due to the highly oscillatory cosine function which is being integrated. Despite $F(\theta)$ is not monotonic in general, its average grows linearly with time before reaching its maximum constant value, when the oscillatory terms compensate themselves. Invariably, the saturation of this function suggests that, in this model, coherence will not be completely washed out for a finite cut-off $\lambda_C$, and will be preserved within a distance
\begin{equation}\label{dC}
d_C\equiv \sigma_0
\left(-\frac{1}{\pi}\frac{d^2I(\theta)}{d\theta^2}\Big|_{\theta=0}\right)^{-1/2},
\end{equation}
which in the high-temperature limit will prove to be of the same order as the de Broglie wavelength of the particle. 

Another way to show this result is by looking at the same time limits of the purity measure (\ref{purity}) which gives
\begin{equation}\label{slimits}
s=\begin{dcases}
1, & \mbox{for } \theta\rightarrow 0,\\
\left(-\frac{I(0)}{\pi^2}\frac{d^2I(\theta)}{d\theta^2}\Big|_{\theta=0}\right)^{-1/2}, & \mbox{for } \theta\rightarrow \infty.
\end{dcases}
\end{equation}

It can be shown that the dominant term of the second derivative of $I(\theta)$ is proportional to $\ln\lambda_C$ (see Appendix \ref{appendix}) for all cases in which $\lambda_C$ is the largest frequency of the problem. For the cases here considered, in which the bath has infinite degrees of freedom, it is valid to assume $\lambda_C\rightarrow \infty$. Therefore, from the above equations, both $d_C$ and $s$ tend to zero with the inverse of $\ln\lambda_C$, showing that the system is slowly led to a diagonal representation of its reduced density matrix, describing a statistical mixture.

\begin{SCfigure}\label{poles}
\includegraphics[width=0.3\textwidth]{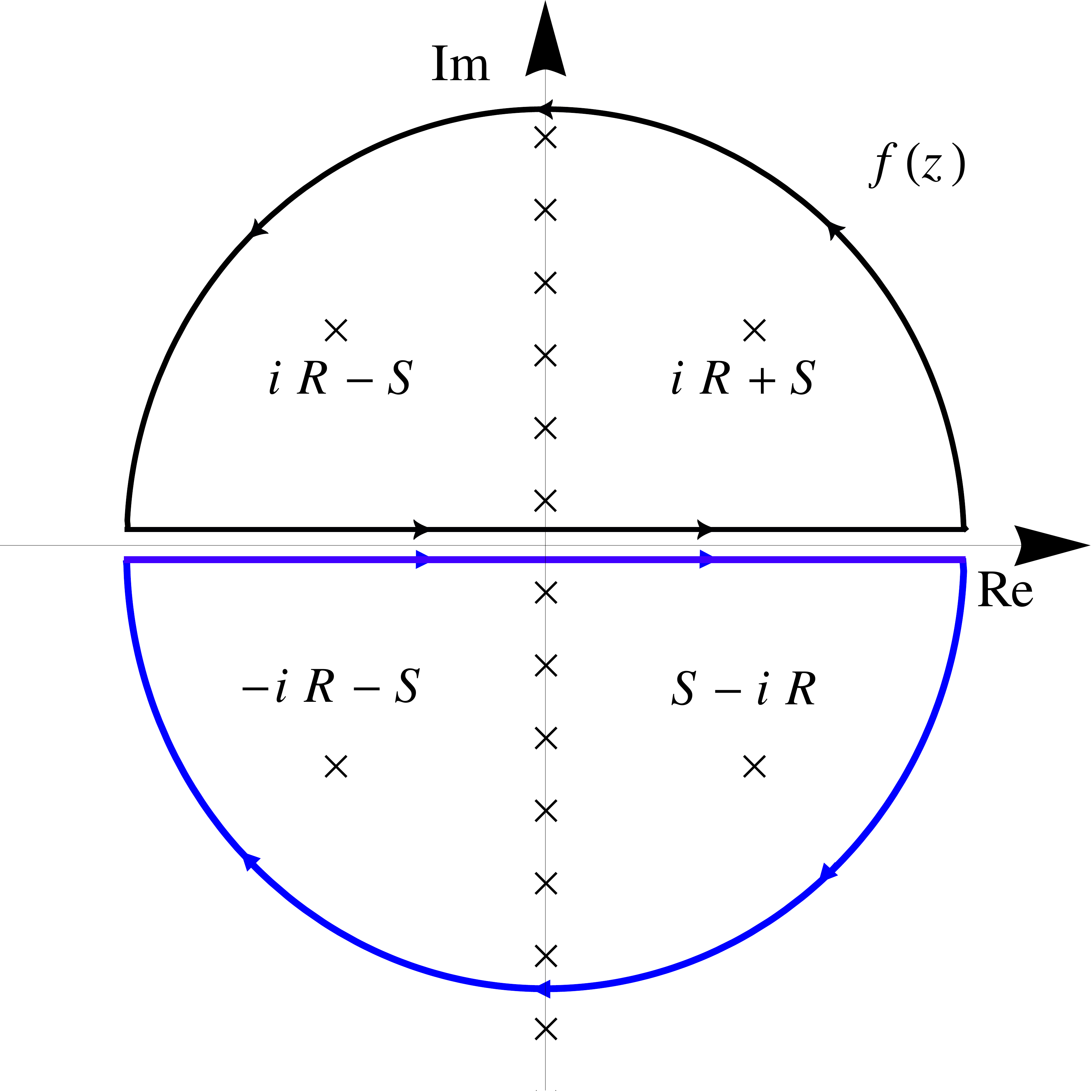} 
\caption{(Color online) Poles of the function $I(\theta)$ and complex paths of integration. When the temperature goes to zero, all the thermal poles that are placed in the imaginary axis go to infinity and thus the residues integration can be performed by neglecting their effects.}
\end{SCfigure}

Now, our analysis of the important time scales can be restricted to the behavior of the last integral. We can verify that the integrand of (\ref{inttheta2-1}) has four symmetric poles, $\lambda_p=\pm(S\pm i R)$, which come from its polynomial denominator and infinite periodic poles, $\lambda_p=i \pi n/\kappa$, with $n$ a non zero integer, from the thermal contribution of the hyperbolic cotangent (see FIG. \ref{poles}). The pole of the latter at $n=0$ gives rise to a removable singularity of the integrand. One should notice that the overdamped limit ($R>1$) can be properly taken into account by making the substitution $S \rightarrow i S$ in all previous expressions. Particularly, this modifies the former four symmetric poles into pure imaginary numbers, $\lambda_p=\pm i(S\pm R)$.

In the case of the underdamped limit, we will always be interested in analyzing time scales that satisfy the conditions $\theta\gg 1$ and $\theta\ll\theta_R$, where $\theta_R$ is the relaxation time of the system. The former guarantees that the exponent (\ref{gamma}) is of order unit for distances $r\approx 1$, allowing us to find the decoherence time scale $\tau_D=\Gamma^{-1}$, as discussed above. The latter assures that we are witnessing dynamical effects on the system, before it thermalizes with the bath. Therefore, what matters to us is the asymptotic behavior of $I(\theta)$ for large times. In the overdamped limit explored in this paper, the first inequality must be substituted by $\theta\ll 1$, once the exponent (\ref{gamma}) in this case is of order one for $r\approx 1$ for very short times. The characteristic time $\theta_R$ assumes the values $\omega_0/\gamma$ and $2\gamma/\omega_0$, in the extremely underdamped and overdamped cases, respectively.

Still in the underdamped regime, since all the time dependence of $I(\theta)$ is in the cosine function, we can generally write the time asymptotic behavior of (\ref{inttheta2-1}) using residues technique as 
\begin{equation}\label{approxit}
I(\theta)\propto \max_{\lambda_p}\left\{e^{-\left|\text{Im}(\lambda_p)\right|\theta}\right\}.
\end{equation}

We see that the frequency $\lambda_p$, associated with one of the poles, that minimizes the value of the exponential dictates the time scale of the decay. Therefore, the temperature dependence of the approximations can be treated separately, in the following limits:

\begin{itemize}
\item \textbf{High-temperature} ($\kappa\rightarrow 0$ ): In this case, assuming that the main contribution to (\ref{inttheta2-1}) comes from frequencies much lower than the thermal frequency, we can replace the hyperbolic cotangent by $1/\kappa\lambda$. This makes the evaluation of $I(\theta)$ straightforward, allowing us to consider only the four remaining poles, for both under and overdamped limits.

\item \textbf{Low-temperature} ($\kappa\rightarrow \infty$): This limit is more subtle. In particular, when one makes $T=0$, this leads to explicit dependence on the cut-off, as we will discuss below. We notice that in this case, the poles of the hyperbolic cotangent cannot be easily neglected, since they go to zero and would in principle have a considerable contribution to (\ref{approxit}). 

However, in the particular situation of extremely weak damping where $R\rightarrow 0$, this low-temperature limit is reasonable if we assume the condition $ R\ll\kappa^{-1}$, so that there is a sufficiently small contribution from the thermal poles. Again, this allows us to ignore their contribution to the large time asymptotic limit. In fact, this conclusion justifies the complete analysis of the weakly damping regime, as developed in Sec.\ref{subweak}.

On the other hand, this last approximation which works for the extremely weak damping scenario is not as useful for the limit $R\rightarrow \infty$. In principle, we can try to extract an analytical result from (\ref{inttheta2-1}) for low-temperatures by considering $\left|\coth(\kappa\lambda)\right|=1$ for all values of $\lambda\neq 0$. The difficulty arises when one needs to evaluate the second derivative of $I(\theta)$ at $\theta=0$, in order to fully obtain the functions (\ref{a1-1}) and (\ref{a3-1}). This leads to a dependence of the resulting functions on $\ln\lambda_C$, and consequently to a logarithmic divergence if one naively forces the frequency cut-off to be infinite (see discussion in the Appendix \ref{appendix}). Due to the cumbersome expressions that come with this limit and to avoid explicit dependence on the cut-off, we will not investigate this limit much further.   

\end{itemize}



\section{Weak Damping}\label{subweak}


As we have discussed, for a weak damping, $R\ll 1$, we are interested in the time dependence of the exponent (\ref{gamma}) satisfying $1\ll\theta\ll R^{-1}$. 

As seen in \cite{Caldeira1989}, if we first compute one of the time integrals in each of the equations (\ref{Athetas1}-\ref{Athetas3}) and then take the limit $R\rightarrow 0$, it can be verified that for $\theta\gg 1$ they have a delta-like behavior, i.e., they are proportional to the Dirac's delta $\delta(\lambda-1)$, centered at the natural frequency $\omega_0$. Therefore, it is straightforward to evaluate (\ref{Athetas}). After that, substituting the resulting expressions and definitions into (\ref{gamma}), we can finally write, for $R\theta\ll 1$,
\begin{equation}
-F(\theta)r^2\approx \phi(\theta)\frac{R\theta}{4}r^2-\frac{r^2}{8\zeta^2},
\end{equation}
where
\begin{equation}
\phi(\theta)\equiv\frac{(\zeta^2-\coth\kappa)\sin^2\theta+\zeta^6\cos^2\theta(1-\zeta^2\coth\kappa)}{(\zeta^4\cos^2\theta+\sin^2\theta)^2}.
\end{equation}

\begin{figure}
\begin{center}
\includegraphics[width=0.5\textwidth]{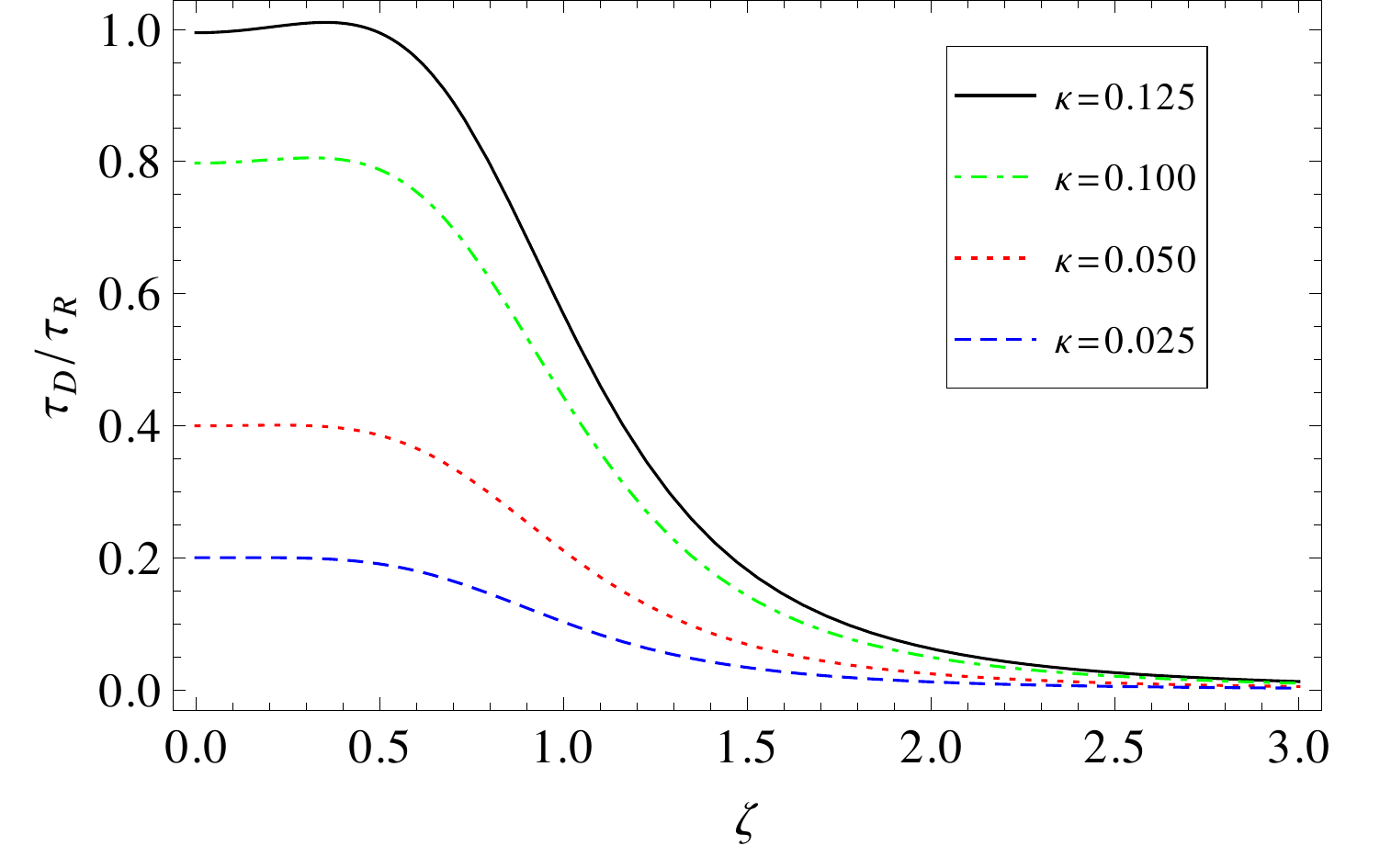} 
\caption{(Color online) Plot of $\tau_D/\tau_R\times\zeta$. We see the dependence of the decoherence time as a function of the squeezing parameter for distances such that $r\approx\zeta$, as we vary the temperature. As the temperature decreases ($\kappa$ increases), the decoherence time becomes longer.}
\label{figure1}
\end{center}
\end{figure}

For a fixed value of both $r$ and $\kappa$, the function $\phi(\theta)$ oscillates with period $\pi$. Therefore, it is reasonable to consider its contribution to the exponent as the average in one period of oscillation. That is
\begin{equation}
\left<\phi(\theta)\right>\equiv\frac{1}{\pi}\int_0^{\pi}\phi(\theta)d\theta=\frac{2\zeta^2-(1+\zeta^4)\coth\kappa}{2\zeta^2}.
\end{equation} 
We then write 
\begin{equation}
\Gamma=-\frac{\gamma}{4}r^2\left<\phi(\theta)\right>
\end{equation}
and so the internal decoherence time is given by
\begin{equation}\label{decweak}
\tau_D\equiv\Gamma^{-1}=\tau_R \left[\frac{8}{(1+\zeta^4)\coth\kappa-2\zeta^2}\right]\left(\frac{\zeta}{r}\right)^{2},
\end{equation}
where $\tau_R\equiv 1/\gamma$ is the relaxation time for the underdamped limit. It is interesting to notice that the decoherence time for distances of order $r\approx \zeta$ is typically shorter than the relaxation time for values of temperature satisfying $0<\kappa\lesssim 0.1$, when the first function inside the brackets in the above equation becomes larger than one (see FIG. \ref{figure1}).

We shall analyze the two extreme temperature limits by making the appropriate approximations to $\coth \kappa$.
\subsection{High-Temperature Limit}
In this case, $\kappa\rightarrow 0$ and $\coth\kappa\approx 1/\kappa$. Obviously, we have the condition $R\ll\kappa^{-1}$ satisfied, which validates the former time scale discussion. Thus, 

\begin{equation}
\tau_D\approx\tau_R\left[\frac{8\kappa}{(1+\zeta^4)-2\kappa \zeta^2}\right]\left(\frac{\zeta}{r}\right)^{2}.
\end{equation}

As expected, we see that the internal decoherence time decreases with $\kappa$.

Due to the particularity of the weak damping limit \cite{Caldeira1989}, we could avoid calculating the explicit form of (\ref{inttheta2-1}) in order to estimate the decoherence time scale $\tau_D$. However, to calculate the residual coherence length $d_C$ given by (\ref{dC}) we must know its explicit form. Again, since $\kappa$ is very small, we can use the high-temperature approximation $\coth\kappa\lambda\approx 1/\kappa\lambda$ in order to solve the integral (\ref{inttheta2-1}) by residues technique. This results in 
\begin{equation}\label{Itempalta1}
I(\theta)\approx\frac{\pi}{\kappa S}e^{-R\theta}[S\cos{(S\theta)}+R\sin{(S\theta)}].
\end{equation}
Inserting (\ref{Itempalta1}) into (\ref{dC}), it follows that
\begin{equation}\label{htdc1}
d_C\equiv
\sigma_0\left(-\frac{1}{\pi}\frac{d^2I(\theta)}{d\theta^2}\Big|_{\theta=0}\right)^{-1/2}=\sigma_0\sqrt{\kappa}=\frac{\lambda_B}{\sqrt{2}},
\end{equation}
in which $\lambda_B\equiv\hbar/\sqrt{2Mk_B T}$ is the thermal de Broglie wavelength, showing that coherence is maintained even in the limit where $\theta\rightarrow \infty$ for distances of order of $\lambda_B$. Consistently, it can be verified that the purity, according to (\ref{slimits}), assumes the asymptotic value 
\begin{equation}\label{purhtc1}
\lim_{\theta\rightarrow\infty}s=\left(-\frac{I(0)}{\pi^2}\frac{d^2I(\theta)}{d\theta^2}\Big|_{\theta=0}\right)^{-1/2}=\kappa\ll 1.
\end{equation}
This temperature dependence of the results was already expected since in this approximation $\kappa$ defines the largest frequency of the problem. 

\subsection{Low-Temperature Limit}
Now, for the low-temperature limit $\kappa$ is very large, implying that $\coth\kappa\approx 1$. Due to the discussion in the last section, it is important to stress the fact that this limit is only consistent keeping $R\ll\kappa^{-1}$, thus prohibiting a zero-temperature analysis. Substituting directly $\coth\kappa\rightarrow 1$ gives us an upper bound to the decoherence time, as  

\begin{equation}\label{tdweak}
\tau_D\approx\tau_R\left(\frac{8}{(1+\zeta^4)-2\zeta^2}\right)\left(\frac{\zeta}{r}\right)^{2}.
\end{equation}

\begin{figure}
\begin{center}
\includegraphics[width=0.5\textwidth]{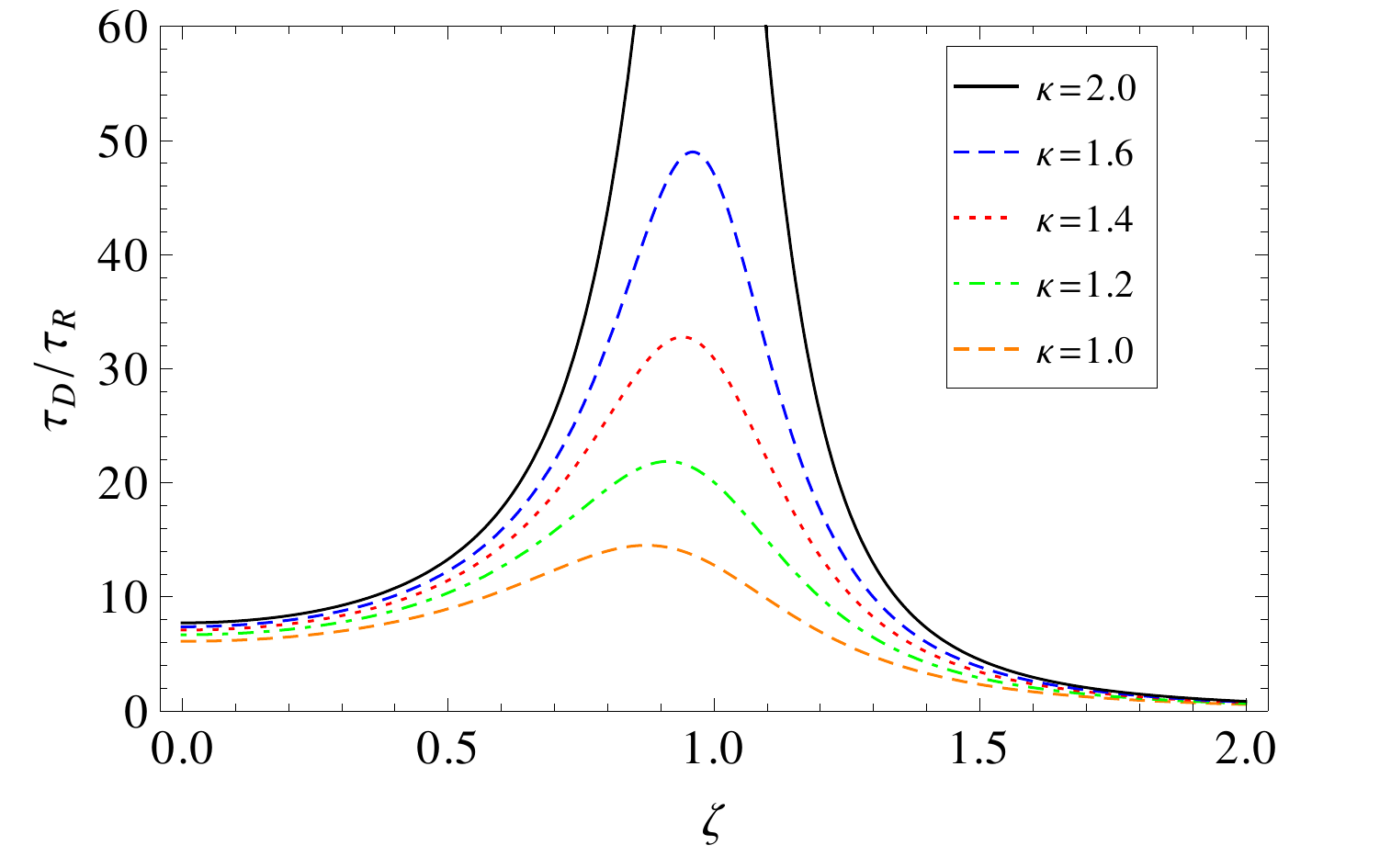} 
\caption{(Color online) Plot of $\tau_D/\tau_R\times\zeta$. Again, assuming  $r\approx\zeta$, in this case of low-temperature, i.e. $\kappa>1$, we see the tendency to the large decoherence time as the value of $\zeta$ approaches one. Clearly, according to (\ref{tdweak}), the decoherence time diverges for zero temperature ($\kappa=\infty$), as suggested by the continuous increase of the peak at $\zeta\approx 1$ as the temperature is decreased.}
\label{figure2}
\end{center}
\end{figure}

We see that when it assumes the value $\zeta=1$ (i.e., no squeezing) the decoherence time $\tau_D$ goes to infinity. This can be understood as a consequence of the fact that in the extremely weak damping limit, the model coincides with the rotating wave approximation \cite{RosenaudaCosta2000}. Furthermore, in \cite{Paz2000} and \cite{Zurek1993}, the authors address the question of how squeezing affects the robustness of a coherent state (which satisfies the minimum Heisenberg's uncertainty) with respect to the decoherence induced by the bath in a harmonic potential under weak dissipation. They show that the ideal squeezed state (with spatial dispersion $\Delta x=\sigma_0$) is the most robust, in agreement with our results. Additionally, from (\ref{tdweak}) it can be seen that the upper-bound value of $\tau_D$ decays drastically for values of $\zeta$ that differs slightly from one, suggesting that the effect of decoherence is fairly sensitive to the initial state (see FIG. \ref{figure2}).

\subsection{Purity}

At last, from (\ref{purity}) we have thus the measurement of purity as a function of time in the limit $\gamma\rightarrow 0$ 
\begin{equation}
s=\text{Tr}\rho^2=1-\frac{\left(\zeta^4+1\right)\coth  \kappa-2\zeta^2}{\zeta^2} \gamma t.
\end{equation}
Consistently, we verify that the trace equals one for $t=0$, as well as the linear decay with time, describing the system losing its purity in the decoherence time scale given by (\ref{decweak}). Obviously, this measurement of purity with linear decay in time holds only for the relevant time scales discussed herein.
\section{Strong Damping}\label{substrong}

In the limit of strong damping, namely $R\rightarrow\infty$, only the case of high-temperature will be studied, following the discussion in the previous section. 

Again, in order to consider the overdamped limit, we must make the substitution $S\rightarrow i S$ in all previous expressions.

\subsection{High-Temperature Limit}

In this case, the analogous expression of (\ref{Itempalta2}) is
\begin{equation}\label{Itempalta2}
I(\theta)\approx\frac{\pi}{\kappa S}e^{-R\theta}[S\cosh{(S\theta)}+R\sinh{(S\theta)}].
\end{equation}
In order to take into account the strong damping limit, we must notice that in this situation the overdamped frequency $S$ has a dependence on $R$ of the form $S\approx R+\mathcal{O}(1/R)$. Due to this fact, after substituting $I(\theta)$ above into equations (\ref{Athetas1}-\ref{Athetas3}), we can expand them for $R\rightarrow\infty$ to first order in 1/R. Afterwards, we consider short times (i.e., $\theta\ll 1$) keeping the product $R\theta$ finite, which allows us to write
\begin{equation}\label{Astrong}
\begin{aligned}
&A^{(1)}(\theta)\approx-\frac{e^{-2R\theta}}{32\kappa}\left(3+4R\theta-4e^{2R\theta}+e^{4R\theta}\right),\\
&A^{(2)}(\theta)\approx\frac{1}{4\kappa}\left[R\theta-\frac{1}{4}\left(e^{2R\theta}-e^{-2R\theta}\right)\right],\\
&A^{(3)}(\theta)\approx\frac{e^{-2R\theta}}{32\kappa}\left[-4+e^{-2R\theta}+e^{2R\theta}(3-4R\theta)\right].
\end{aligned}
\end{equation}
Substituting (\ref{Astrong}) into (\ref{gamma}) leaves us with
\begin{equation}\label{exp1}
F(\theta)\approx\frac{(1-e^{-4R\theta})\zeta^2+\kappa e^{-4R\theta}}{8\kappa\zeta^2}.
\end{equation}
We can then expand the expression above for $R\theta\ll 1$ in order to extract its linear behavior as shown in  
\begin{equation}\label{exp}
-F(\theta)r^2\approx -\frac{R\theta}{2}r^2\left(\frac{\zeta^2-\kappa}{\kappa\zeta^2}\right)-\frac{r^2}{8\zeta^2}.
\end{equation}
Consequently, the decay rate 
\begin{equation}
\Gamma=\frac{\gamma}{2}r^2\left(\frac{\zeta^2-\kappa}{\kappa\zeta^2}\right),
\end{equation}
makes the internal decoherence time be simply given by
\begin{equation}\label{td}
\tau_D\equiv\Gamma^{-1}=\frac{2}{\gamma}\left(\frac{\kappa}{\zeta^2-\kappa}\right)\left(\frac{\zeta}{r}\right)^2.
\end{equation}
Once the relaxation time for an overdamped harmonic oscillator is given by $\tau_R=2\gamma/\omega_0^2$ , we can rewrite (\ref{td}) as
\begin{equation}\label{td2}
\tau_D=\tau_R\left[\frac{1}{R^2}\left(\frac{\kappa}{\zeta^2-\kappa}\right)\right]\left(\frac{\zeta}{r}\right)^2.
\end{equation}
Since $\kappa\ll 1$ and $R\rightarrow \infty$, this shows that the decoherence time for the strong damping and high temperature limit is typically much shorter than the relaxation time, once the expression inside the brackets becomes very small.
 
Finally, following the same expansion for $R\theta\ll 1$ in (\ref{purity}), we have for the purity 
\begin{equation}
s=\text{Tr}\rho^2=1-2\left(\frac{\zeta^2 }{\kappa}-1\right)\gamma t\approx 1-2 \frac{\zeta^2 }{\kappa}\gamma t.
\end{equation}

Inserting (\ref{Itempalta2}) into (\ref{dC}), it follows that
\begin{equation}\label{htdc}
d_C\equiv
\sigma_0\left(-\frac{1}{\pi}\frac{d^2I(\theta)}{d\theta^2}\Big|_{\theta=0}\right)^{-1/2}=\sigma_0\sqrt{\kappa}=\frac{\lambda_B}{\sqrt{2}}.
\end{equation}
Analogously, according to (\ref{slimits}), the purity assumes the asymptotic value 
\begin{equation}\label{purhtc}
\lim_{\theta\rightarrow\infty}s=\left(-\frac{I(0)}{\pi^2}\frac{d^2I(\theta)}{d\theta^2}\Big|_{\theta=0}\right)^{-1/2}=\kappa\ll 1.
\end{equation}
As expected, both (\ref{htdc}) and (\ref{purhtc}) are in agreement with the previous results for the case of weak damping, since we have already anticipated their behavior solely in the high-temperature approximation. 

For $\zeta^2>\kappa$ and $\zeta^2<\kappa$, we verify that the growth of the exponent given by (\ref{exp}), which is associated with the off-diagonal terms of (\ref{roabrev}), can be respectively negative and positive. For this limit, the final coherence of the packet should be always the same within a length of the order of $\lambda_B$, due to (\ref{htdc}). Consequently, a positive growth implies that the packet would gain coherence from the coupling with the bath, meaning that the off-diagonal terms of (\ref{roabrev}) grow until the final coherence width $d_C$ is reached. It should be noticed that the final coherence length is negligible, as a consequence of the high-temperature approximation. Additionally, this is even more prominent for macroscopic systems, in which the mass $M$ is large, making $\lambda_B\rightarrow 0$. This same high-temperature analysis naturally holds for the underdamped limit. 

\section{Discussion}\label{discussion}

We have analytically estimated the internal decoherence time for a single Gaussian wave packet, as a function of its initial width and temperature, defined by the thermal equilibrium state with a bath of harmonic oscillators. 

For extremely weak damping we could make estimations for the internal decoherence time for both limits of temperature without the explicit dependence on the cut-off frequency. The conclusions for the strong damping case showed to be less straightforward and more restrictive, ruling out a generalization as in the previous case. Nevertheless, for both cases we could verify that in the high-temperature limit some coherence is still left within a length scale proportional to the thermal de Broglie wavelength, even after thermalization. Moreover, for the low-temperature limit, the coherence length depends explicitly on the cut-off frequency as $\ln\lambda_C$. Accordingly, the particle loses its coherence but does not become completely classical, in the sense that the reduced density matrix still maintains an off-diagonal width. At any rate, this approach is sufficient to suggest a tendency towards classicality of the single Gaussian state as it interacts with a large bath.

Particularly, our results agree with references \cite{Venugopalan1994} and \cite{Hakim1985}, which treat the problem of a free particle in high and low-temperature limits, respectively. In the limit of strong damping our problem is analogous to that of a free particle, since in this case the harmonic oscillator frequency can be neglected. Reference \cite{Venugopalan1994} has as well verified the thermal de Broglie wavelength as a residual coherence length, whereas in reference \cite{Hakim1985} the authors have calculated the squared momentum expectation value, which carries information on the off-diagonal terms, and also verified the above mentioned logarithmic dependence on the cut-off.  

We could additionally confirm the robustness of the ideal squeezed state (i.e., $\zeta=1$) compared to other squeezed initial states, as asserted previously by other studies \cite{Zurek1993}. For this special model, we verify that squeezing will not smooth out the effects of decoherence, on the contrary, the system tends to lose internal coherence even faster. Therefore, its worth noticing the importance of the initial state preparation to this typical time scale.

\appendix*
\section{Divergence for the zero-temperature limit and strong dissipation}\label{appendix}

The integration of (\ref{inttheta2-1}) can be made by the use of residues technique, with some caution. Evaluating the limit,
\begin{equation}\label{intthetaT0}
\begin{aligned}
I(\theta)\equiv&\lim_{\kappa\rightarrow \infty}\frac{1}{2}\int_{-\lambda_C}^{\lambda_C}d\lambda\frac{4R\lambda}{(\lambda^2-1)^2+4R^2\lambda^2}\\
&\times\coth(\kappa\lambda)\cos(\lambda\theta).
\end{aligned}
\end{equation}
we can write
\begin{equation}\label{A1}
I(\theta)\approx\int_{0}^{\lambda_C}d\lambda\frac{4R\lambda\cos(\lambda\theta)}{(\lambda^2-1)^2+4R^2\lambda^2},
\end{equation}
where we have assumed the validity of $\coth{\kappa\lambda}\rightarrow 1$. By doing this, we are eliminating an infinite number of imaginary poles due to the hyperbolic cotangent term. Since we are interested in the limit $R\rightarrow \infty$, their contributions can, in principle, be neglected in comparison with the four remaining poles.

Using the fact that in the overdamped regime we have $R^2-S^2=1$, the four remaining poles of (\ref{A1}) are then imaginary, and we can write
\begin{equation}\label{A2}
\begin{aligned}
I(\theta)&\approx\frac{1}{2 S}\int_{0}^{\infty}d\lambda\Bigg\{\frac{1}{\lambda+i(R-S)}-\frac{1}{\lambda-i(R+S)}\\
&-\frac{1}{\lambda+i(R+S)}+\frac{1}{\lambda-i(R-S)}\Bigg\}\cos(\lambda\theta).
\end{aligned}
\end{equation}

Each integral contributes, respectively, with 
\begin{equation}
\begin{aligned}
&\int_{0}^{\infty}d\lambda\frac{\cos(\lambda\theta)}{\lambda\pm i(R\mp S)}=\mp i\pi e^{-(R\pm S)\theta}+\\
&\int_{0}^{\infty}dx \left[\frac{1}{x-(R\mp S)}\pm\frac{1}{x+(R\mp S)}\right]e^{-\theta x},
\end{aligned}
\end{equation}

\begin{equation}
\begin{aligned}
&\int_{0}^{\infty}d\lambda\frac{\cos(\lambda\theta)}{\lambda\pm i(R\pm S)}=\mp i\pi e^{-(R\pm S)\theta}+\\
&\int_{0}^{\infty}dx \left[\frac{1}{x-(R\pm S)}+\frac{1}{x+(R\pm S)}\right]e^{-\theta x},
\end{aligned}
\end{equation}
where the imaginary parts come from the half contributions from the residues of the poles and the real parts from integrations over the imaginary axis, which were evaluated as principal valued. Summing up the repeated terms we can finally write (\ref{A1}) according to 
\begin{equation}\label{correct}
\begin{aligned}
I(\theta)&\approx e^{R\theta}\int_{R}^{\infty}dx \frac{e^{-\theta x}}{x^2-R^2+1}\\
&-e^{-R\theta}\int_{-R}^{\infty}dx \frac{e^{-\theta x}}{x^2-R^2+1},
\end{aligned}
\end{equation} 
which can be rewritten in terms of the special functions gamma and exponential integral as follows 
\begin{equation}\label{correct2}
\begin{aligned}
I(\theta)&\approx \frac{1}{2R}\Bigg\{e^{\frac{\theta}{2 R}}\Gamma\left(0,\frac{\theta}{2 R}\right)-e^{2 R\theta}\Gamma\left(0,2 R\theta\right)\\
&+e^{-2 R\theta}E_i\left(2 R\theta\right)-e^{-\frac{\theta}{2 R}}E_i\left(\frac{\theta}{2 R}\right)\Bigg\}.
\end{aligned}
\end{equation} 
At the strong damping ($R\rightarrow \infty$) limit, the exponential integrals are dominant and from their contributions, it can be shown that the second derivative of $I(\theta)$ is divergent.

Another straightforward way to verify this issue is by analyzing (\ref{A1}) at the limit $\lambda_C\rightarrow \infty$. We can verify the existence of a logarithmic divergence in the evaluation of
\begin{equation}\label{div}
\begin{aligned}
\frac{d^2I(\theta)}{d\theta^2}\Bigg|_{\theta=0}&\approx-\int_{0}^{\lambda_C}d\lambda\frac{4R\lambda^3}{(\lambda^2-1)^2+4R^2\lambda^2}\\
&\propto \ln\lambda_C\rightarrow \infty,
\end{aligned}
\end{equation}
which is not compensated by any other term in the exponent (\ref{gamma}). This qualitative result holds for any case in which $\lambda_C$ is the highest frequency considered, since $\coth(\kappa\lambda)\rightarrow 1$ for $\lambda\rightarrow \infty$ and inevitably the logarithmic divergence at the cut-off appears.

\begin{acknowledgments}
Financial support by Funda\c{c}\~{a}o de Amparo \`{a} Pesquisa do Estado de S\~{a}o Paulo (FAPESP/Brazil), National Institute for Science and
Technology of Quantum Information (INCT-IQ), and Conselho Nacional de Pesquisa (CNPq/Brazil) are gratefully acknowledged.
\end{acknowledgments}



\bibliography{bib}

\end{document}